\documentclass[lettersize,journal]{IEEEtran}
\usepackage{amsmath,amsfonts}
\usepackage{algorithmic}
\usepackage{algorithm}
\usepackage{array}
\usepackage[caption=false,font=normalsize,labelfont=sf,textfont=sf]{subfig}
\usepackage{textcomp}
\usepackage{stfloats}
\usepackage{url}
\usepackage{verbatim}
\usepackage{graphicx}
\usepackage{cite}
\usepackage{amsmath}
\usepackage{amssymb}
\usepackage{booktabs}
\usepackage{makecell}
\usepackage{multirow}
\usepackage{xcolor}
\usepackage[table]{xcolor}
\usepackage{colortbl}

\definecolor{lowred}{RGB}{246, 222, 222}      
\definecolor{midpurple}{RGB}{225, 220, 243}   
\definecolor{highgreen}{RGB}{220, 239, 220}   

\newcommand{\scorecell}[1]{%
  \ifdim #1pt < 0.5pt
    \cellcolor{lowred}#1%
  \else
    \ifdim #1pt < 0.8pt
      \cellcolor{midpurple}#1%
    \else
      \cellcolor{highgreen}#1%
    \fi
  \fi
}
\begin{document}

\title{From Representations to Computational Trajectories: Intrinsic Fingerprinting for Large Language Model Provenance Verification}

\author{Haiyong Tang, Hanzhou Wu, Xinpeng Zhang, and Athanasios V. Vasilakos\thanks{Hanzhou Wu and Xinpeng Zhang are the corresponding authors.}\thanks{Haiyong Tang, Hanzhou Wu, and Xinpeng Zhang are with the School of Communication and Information Engineering, Shanghai University, Shanghai 200444, China (email: haiyongtang@shu.edu.cn; h.wu.phd@ieee.org; xzhang@shu.edu.cn).}
\thanks{Athanasios V. Vasilakos is with the Department of ICT and Center for AI Research, University of Agder (UiA), Jon Lilletuns vei 9, 4879 Grimstad, Norway (email: th.vasilakos@gmail.com).}}

      



\maketitle

\begin{abstract}
Large language models (LLMs) are valuable digital assets, yet their weights can be easily modified after unauthorized acquisition, making model provenance difficult to verify. Model fingerprinting provides a promising solution for model ownership protection without modifying the model or embedding ownership information. However, parameter-based fingerprints can be sensitive to model transformations, limiting their robustness, while representation-based fingerprints may assign high similarity to unrelated models, limiting their uniqueness. To address these limitations, we propose TrFP, an intrinsic fingerprinting method based on computational trajectories. Instead of comparing model parameters or representations, TrFP captures how representations evolve across network depth and constructs compact fingerprints using permutation-invariant trajectory descriptors. Experiments across multiple LLM families show that TrFP reliably distinguishes related and unrelated models. It also remains robust under fine-tuning, pruning, quantization, model merging, parameter permutation, and adaptive attacks, while requiring less than 0.3 MiB of fingerprint storage. Overall, TrFP provides stronger provenance discrimination than several existing state-of-the-art methods, demonstrating the effectiveness of computational trajectories for reliable LLM provenance verification.  
\end{abstract}

\begin{IEEEkeywords}
Large language models, model provenance verification, model fingerprinting, security, ownership verification.
\end{IEEEkeywords}

\section{Introduction}
Developing large language models (LLMs) requires extensive data, computing resources, and engineering effort.
As a result, trained LLMs are valuable digital assets with significant intellectual property and commercial value~\cite{TIFS_watermark1,yan2025enhancing,shao2026fit}. 
Once model weights are obtained without authorization, attackers can easily modify them using common techniques such as fine-tuning~\cite{Chiang2023Vicuna}, pruning~\cite{Xia2024ShearedLLaMA}, quantization~\cite{frantar2022gptq}, or parameter permutation~\cite{entezari2021role}.
These modifications can preserve the main capabilities of the model while hiding its relationship with the original model. 
This may enable unauthorized use or commercialization and cause economic losses and security risks to model owners~\cite{lederer2023identifying}.
Therefore, reliably tracing the provenance of suspicious models has become an important problem in LLM security~\cite{yang2024fingerprint, zhao2026robust}.

Existing model provenance verification techniques can be broadly divided into watermarking and fingerprinting.
Model watermarking~\cite{masrani2025task,yang2026codegenguard} injects ownership information into model parameters or behaviors before model release.
However, such methods may affect model performance and cannot protect models that have already been released.
In contrast, model fingerprinting~\cite{Zeng2024HuRef,zhang2025self,wang2025ghost,zeng2026awm,zhang2024easydetector,Zhang2025REEF,liu2026fnf,tang2026cref,zhang2026attndiff} uses the model's inherent features for provenance verification and requires no pre-embedded ownership information.
In general, an effective fingerprint should satisfy two essential properties, namely, robustness and uniqueness.
Robustness means that a model derived from the source model should retain high fingerprint similarity after model transformations. Uniqueness means that a model unrelated to the source model should yield low fingerprint similarity, even when it shares the same architecture.
Based on the features used, fingerprinting methods can be further divided into parameter-based and representation-based methods.
Parameter-based methods~\cite{Zeng2024HuRef,zhang2025self,wang2025ghost,zeng2026awm} construct fingerprints from model parameters and achieve strong uniqueness when parameter shapes and correspondence are preserved. 
However, extensive continued training and structured pruning can substantially change parameter features, which limits their robustness.
Representation-based methods~\cite{zhang2024easydetector,Zhang2025REEF,liu2026fnf,tang2026cref,zhang2026attndiff} construct fingerprints from the internal representations, reducing the dependence on direct parameter correspondence.
Since hidden dimensions may not strictly correspond across models, these methods commonly use Centered Kernel Alignment (CKA)~\cite{CKA} to measure representation similarity and treat high similarity as evidence of a provenance relationship.

\begin{figure}[t]
  \centering
  \includegraphics[width=\linewidth]{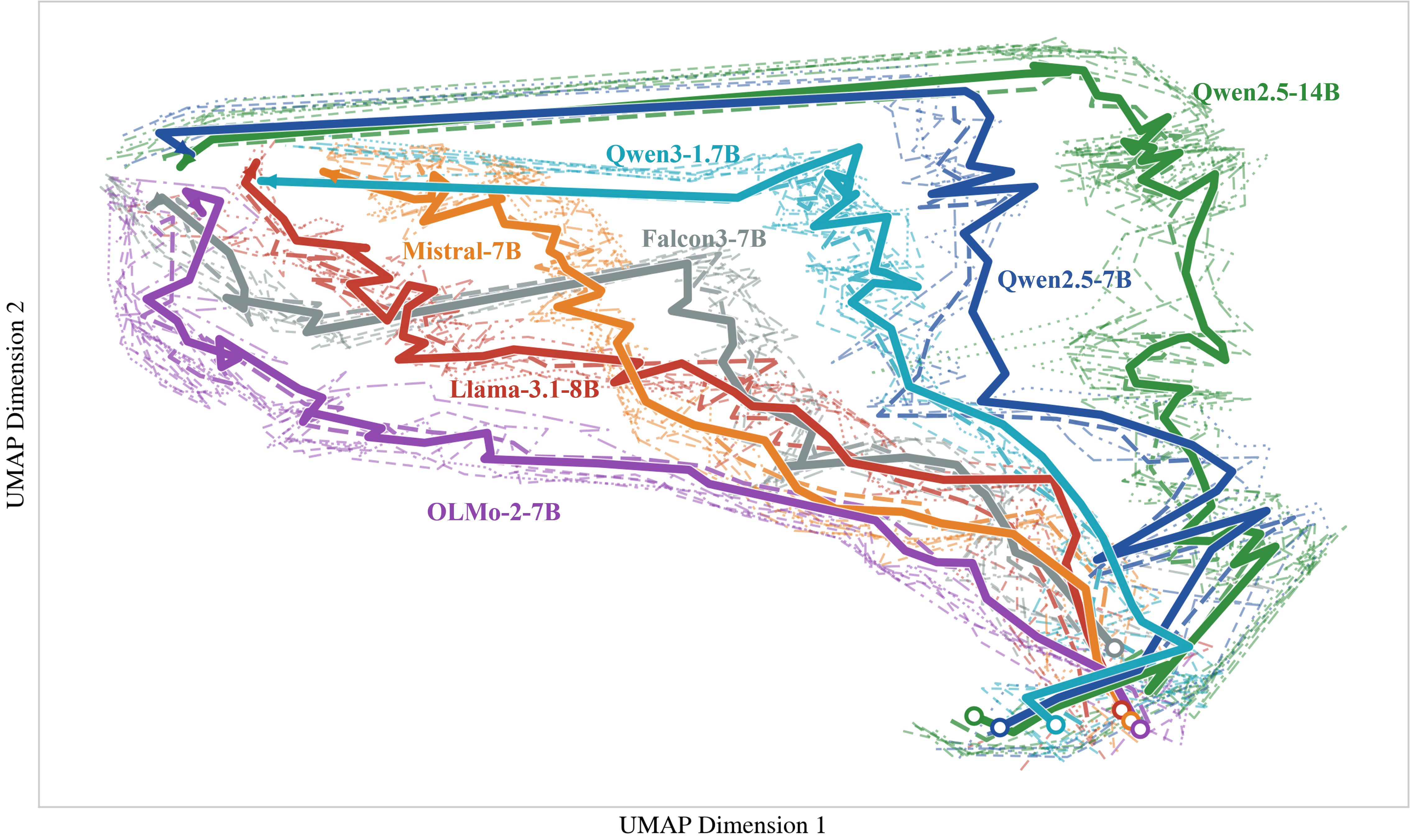}
  \vspace{-20pt}
  \caption{Computational trajectories of different LLM families and their derived models. Different model families follow distinct trajectories, whereas derived models stay close to the trajectories of their source models.}
  \label{fig0_motivation}
  \vspace{-10pt}
\end{figure}

\subsection{Motivation}
However, using high representation similarity as evidence of model provenance requires two conditions. 
(1) The similarity metric should reliably measure representation similarity. 
(2) Representation similarity should be specific to model provenance. 
We find that neither condition is always satisfied in existing representation-based fingerprinting. 
CKA can substantially overestimate similarity under high-dimensional, small-sample settings, while unrelated models can still exhibit high similarity. Together, these issues undermine the uniqueness of representation-based fingerprints.
In summary, parameter-based fingerprints suffer from limited robustness to model transformations, whereas representation-based fingerprints lack sufficient uniqueness. This motivates us to seek information that remains stable under model transformations while distinguishing unrelated models. Therefore, we examine how representations evolve across layers rather than only the representations themselves.
In Transformer-based LLMs, each layer updates the hidden state according to its parameters, so models with similar representations need not follow the same layer-wise update process.
We define this sequence of layer-wise updates along network depth as the computational trajectory.
As shown in Fig.~\ref{fig0_motivation}, the computational trajectories of different model families are clearly separated, whereas derived models largely follow the trajectories of their source models.
This suggests that computational trajectories can provide both uniqueness and robustness.
First, although representations from different models may converge, their layer-wise updates are determined by their respective parameters~\cite{elhage2021mathematical}, allowing the computational trajectories of unrelated models to remain distinguishable.
Second, derived models are modified from their source models rather than trained independently from scratch~\cite{Zeng2024HuRef,Zhang2025REEF}, allowing their computational trajectories to retain characteristics inherited from the source model.

Computational trajectories provide a strong basis for provenance verification. However, comparing trajectories across models is nontrivial because their hidden spaces have no coordinate correspondence~\cite{raghu2017svcca,CKA}. Such comparison requires characterizing trajectories without relying on hidden coordinates. To this end, we propose TrFP, which compares models not by where their trajectories lie in hidden spaces, but by how the trajectories evolve across network depth.
TrFP characterizes this evolution by how strongly each layer updates the representation, how this update changes between neighboring layers, and how far the representation has moved from its starting point.
As a result, the fingerprint is invariant to parameter permutation by construction, requires no parameter correspondence, and remains compact as model size grows.

\subsection{Main Contributions}
In summary, the main contributions of this paper include:
\begin{itemize}
    \item We reveal a fundamental limitation of existing representation-based LLM fingerprinting methods. We show that representation similarity can be overestimated due to estimation bias, while unrelated models can still exhibit high similarity due to representation convergence.
    \item We introduce computational trajectories as a new perspective and propose TrFP. It captures how representations evolve across network depth using compact trajectory descriptors, achieving stronger uniqueness and robustness than several existing state-of-the-art methods.
    \item We systematically evaluate TrFP under various model transformations and attacks. TrFP effectively distinguishes related and unrelated models and remains robust to fine-tuning, pruning, quantization, model merging, parameter permutation, and adaptive attacks, with a fingerprint size below 0.3 MiB.
\end{itemize}

\section{Related Works}
\subsection{Deep Neural Network Fingerprinting}
Early deep neural network (DNN) fingerprinting methods mainly characterize models through their decision behaviors. IPGuard~\cite{cao2021ipguard} extracts fingerprint samples near the classification boundary, based on the observation that pirated models tend to inherit the decision boundary of the source model. Since adversarial examples near the boundary may also transfer to independently trained models, Lukas et al.~\cite{lukas2021deep} construct conferrable adversarial examples that transfer only to surrogate models. Peng et al.~\cite{peng2022fingerprinting} further exploit universal adversarial perturbations to characterize the global geometry of model decision boundaries.
Beyond model behaviors, model parameters provide intrinsic characteristics that do not depend on crafted queries. Zheng et al.~\cite{zheng2022dnn} generate compact fingerprints by projecting distinctive model weights into a random space determined by the owner. Recent studies further focus on the robustness and security of model fingerprints, since attackers may modify the model or even forge ownership claims. GMFIP~\cite{yan2025enhancing} trains a generator to synthesize fingerprint samples, improving robustness against model modifications and extraction attacks. FIT-Print~\cite{shao2026fit} reveals the vulnerability of conventional fingerprints to false ownership claims and introduces targeted fingerprints for more reliable ownership verification. Overall, these studies establish the foundation for fingerprinting large language models.

\begin{figure*}[t]
  \centering
  \includegraphics[width=\linewidth]{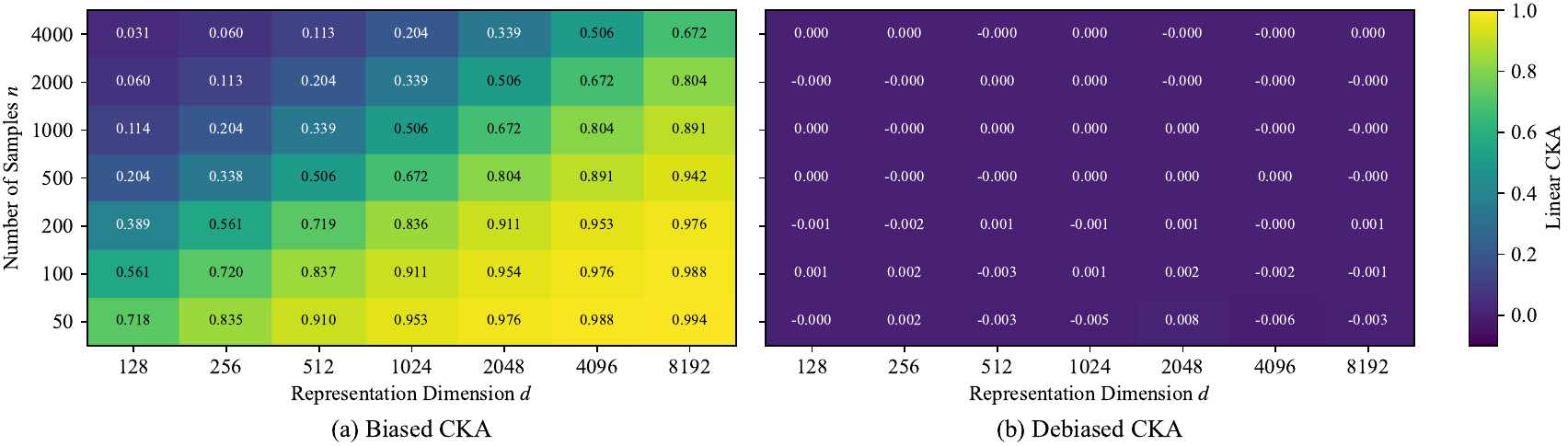}
     \vspace{-20pt}
  \caption{Linear CKA between two unrelated Gaussian random representations under different numbers of samples $n$ and representation dimensions $d$. }
  \label{fig2:random_represent}
  \vspace{-10pt}
\end{figure*}

\subsection{Large Language Model Fingerprinting}
Unlike conventional DNNs, many LLMs are released with open weights and adapted into derived models, which makes provenance verification with white-box access both practical and necessary. 
Early studies mainly construct fingerprints from model parameters.  
PCS~\cite{Zeng2024HuRef} finds that parameter directions remain stable after pre-training and uses cosine similarity between parameter vectors to determine model provenance. 
ICS~\cite{Zeng2024HuRef} further exploits parameter invariants in Transformer attention and feed-forward modules to improve robustness against parameter permutation. 
Recent studies explore parameter structures that are more stable under model transformations. 
SELF~\cite{zhang2025self} constructs invariant matrices from attention weights and uses their singular values and eigenvalues as fingerprints. GhostSpec~\cite{wang2025ghost} extracts compact fingerprints from the spectral structure of attention weight matrices and uses layer alignment to compare models with different depths. 
AWM~\cite{zeng2026awm} instead treats weight comparison as a parameter alignment problem, aligning parameter coordinates before measuring their similarity. 
However, these methods rely on parameter features remaining stable and comparable after model transformations, which may limit their robustness to large parameter updates, structural changes, or weak parameter correspondence.

To reduce the dependence on direct parameter correspondence, recent studies have explored internal representations for model fingerprinting.
The basic idea is that models derived from the same source may retain similar representation features after model transformations. 
EasyDetector~\cite{zhang2024easydetector} trains a classifier using intermediate representations and determines model provenance based on how well the classifier transfers to the suspect model. 
REEF~\cite{Zhang2025REEF} directly compares intermediate representations produced by two models on the same inputs and uses CKA to measure their similarity without requiring strict correspondence between hidden dimensions. 
FNF~\cite{liu2026fnf} identifies groups of neurons that are activated together and compares their activation trends on the same inputs. CREF~\cite{tang2026cref} constructs semantic concept directions in the representation space and uses model responses along these directions as compact fingerprints.
AttnDiff~\cite{zhang2026attndiff} constructs semantically conflicting prompt pairs and uses the singular values of attention-map differences as fingerprints.

Overall, representation-based methods reduce the dependence on direct parameter correspondence by exploiting internal model responses. 
However, these methods rely on representation similarity as evidence of shared provenance.
Such similarity may not be specific to provenance relationships, potentially limiting the uniqueness of the resulting fingerprints. We systematically examine this issue in the next section.

\section{Revisiting Representation Similarity for Model Fingerprinting}
Existing representation-based methods typically determine model provenance by measuring representation similarity between a source model and a suspect model. For representation similarity to reliably indicate model provenance, two conditions should hold: (1) the similarity metric should reliably measure representation similarity, and (2) representation similarity should be specific to model provenance. In this section, we examine whether these two conditions hold and analyze their implications for existing representation-based methods.

\subsection{Reliability of Representation Similarity Measurement}
Recent representation-based methods, such as REEF~\cite{Zhang2025REEF}, CREF~\cite{tang2026cref}, and AttnDiff~\cite{zhang2026attndiff}, measure representation similarity with CKA~\cite{CKA}, because it is invariant to permutations of hidden dimensions and requires no coordinate correspondence across models. Therefore, we focus on whether CKA can reliably measure representation similarity.

Let $\mathbf{X} \in \mathbb{R}^{n \times d_1}$ and $\mathbf{Y} \in \mathbb{R}^{n \times d_2}$ be two representation matrices, where $n$ denotes the number of samples, and $d_1$ and $d_2$ denote the representation dimensions.
Linear CKA constructs the Gram matrices  $\mathbf{K}=\mathbf{X}\mathbf{X}^\top$ and $\mathbf{L}=\mathbf{Y}\mathbf{Y}^\top$ and measures their dependence using the Hilbert-Schmidt Independence Criterion (HSIC)~\cite{gretton2005measuring}:
\begin{align}
\mathrm{CKA}(\mathbf{K}, \mathbf{L}) &=
\frac{\mathrm{HSIC}(\mathbf{K}, \mathbf{L})}
{\sqrt{\mathrm{HSIC}(\mathbf{K}, \mathbf{K})\,\mathrm{HSIC}(\mathbf{L}, \mathbf{L})}}, 
\label{eq:CKA}\\
\mathrm{HSIC}(\mathbf{K}, \mathbf{L}) &=
\frac{1}{(n-1)^2}\,\mathrm{tr}(\mathbf{K}\mathbf{H}\mathbf{L}\mathbf{H}), 
\end{align}
where $\mathbf{H} = \mathbf{I}_n - \frac{1}{n}\mathbf{1}\mathbf{1}^{\top}$ is the centering matrix, $\mathbf{I}_n$ is the $n \times n$ identity matrix, and $\mathbf{1} \in \mathbb{R}^{n}$ is the all-ones vector.
In model fingerprinting, a high CKA score is commonly viewed as evidence that two models have similar representations. 
Accordingly, unrelated representations are expected to yield low CKA scores. 

To examine whether this expected behavior holds, we generate two unrelated Gaussian random representations. 
We then compute their CKA scores under different numbers of samples $n$ and representation dimensions $d$. 
Since the two representations are generated independently, no underlying correspondence exists between them. 
The results are shown in Fig.~\ref{fig2:random_represent}(a).
CKA varies substantially with both $n$ and $d$. 
When $d$ is much larger than $n$, CKA between unrelated random representations can approach 1. 
For example, CKA reaches 0.976 when $n=200$ and $d=8192$.
Such high-dimensional, small-sample settings are common in LLM fingerprinting. For example, REEF~\cite{Zhang2025REEF} uses 200 probe samples to compute CKA between model representations. 

We trace this phenomenon to the finite sample bias of the biased HSIC estimator~\cite{murphy2024correcting}, which existing representation-based fingerprints inherit.
When $d \gg n$, high-dimensional random vectors tend to be nearly orthogonal. 
Their Gram matrices are then dominated by self-similarity terms on the diagonal. 
As a result, unrelated representations may still obtain high CKA scores.
To examine whether this bias can be corrected, we use the unbiased HSIC estimator~\cite{song2012feature}:
\begin{equation}
\begin{split}
\mathrm{HSIC}_u(\mathbf{K}, \mathbf{L}) = \frac{1}{n(n-3)} \Bigg[\,
& \mathrm{tr}(\tilde{\mathbf{K}}\tilde{\mathbf{L}})
+ \frac{\mathbf{1}^{\top}\tilde{\mathbf{K}}\mathbf{1}\,\mathbf{1}^{\top}\tilde{\mathbf{L}}\mathbf{1}}{(n-1)(n-2)} \\
& - \frac{2}{n-2}\,\mathbf{1}^{\top}\tilde{\mathbf{K}}\tilde{\mathbf{L}}\mathbf{1}
\,\Bigg],
\end{split}
\label{eq:hsic_u}
\end{equation}
where $\tilde{\mathbf{K}}$ and $\tilde{\mathbf{L}}$ are the Gram matrices with their diagonals set to zero. 
Replacing HSIC in Eq.~\eqref{eq:CKA} with $\mathrm{HSIC}_u$ gives debiased CKA. As shown in Fig.~\ref{fig2:random_represent}(b), the high similarity between unrelated random representations largely disappears after debiasing. 
For example, when $n=200$ and $d=8192$, the CKA score decreases from 0.976 to only 0.001 after debiasing.

These results show that CKA can substantially overestimate representation similarity under high-dimensional, small-sample settings due to estimation bias. However, this bias can be effectively corrected using debiased CKA. 

\begin{figure*}[t]
  \centering
  \includegraphics[width=\linewidth]{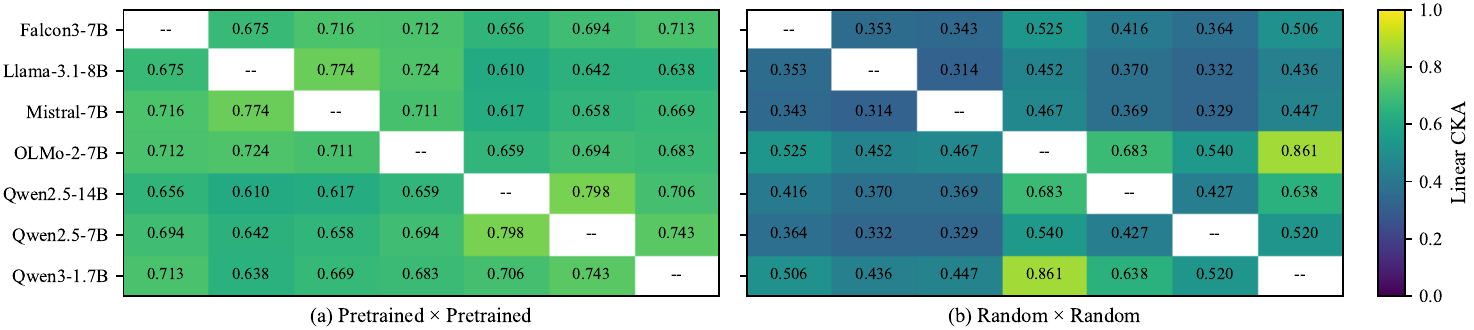}
   \vspace{-20pt}
  \caption{Pairwise debiased CKA between representations of different models
before and after pretraining. Rows follow the same model order as the columns.}
  \label{fig3:pretrained_vs_random}
   \vspace{-10pt}
\end{figure*}

\subsection{Provenance Specificity of Representation Similarity}
Correcting the estimation bias improves the reliability of representation similarity measurement, but does not establish its specificity to model provenance. If unrelated models can also exhibit high representation similarity, such similarity cannot reliably indicate a provenance relationship.
To examine this question, we select seven pretrained models from several widely used open-source LLM families: Falcon3-7b~\cite{Falcon3}, Llama-3.1-8b~\cite{dubey2024llama3}, Mistral-7b~\cite{jiang2023mistral}, OLMo-2-7b~\cite{olmo2024olmo2}, Qwen2.5-14b~\cite{yang2024qwen25}, Qwen2.5-7b~\cite{yang2024qwen25}, and Qwen3-1.7b~\cite{yang2025qwen3}. 
These models have no derivation relationships with each other. 
We randomly select 200 samples from TruthfulQA~\cite{Lin2022TruthfulQA} as the probe set. 
For each input, we extract the hidden state of the last token at each layer and compute debiased CKA.
The results are shown in Fig.~\ref{fig3:pretrained_vs_random}(a). 
Even after removing the estimation bias, unrelated models still exhibit high representation similarity.
The similarities of all model pairs range from 0.61 to 0.80, with an average of 0.69. 
For example, Llama-3.1-8b and Mistral-7b achieve 0.774, while Qwen2.5-7b and Qwen2.5-14b achieve 0.798.
For reference, REEF~\cite{Zhang2025REEF} regards CKA scores below 0.5 as indicating unrelated models, while all model pairs here exceed this threshold.

One possible explanation is that the high similarity mainly comes from the shared Transformer architecture.
To examine this possibility, we retain the architectures of the seven models but randomly reinitialize all their parameters. 
These models contain no knowledge learned during training. 
We then compute debiased CKA using the same probe set and experimental settings.
As shown in Fig.~\ref{fig3:pretrained_vs_random}(b), the similarities between randomly initialized models are much lower than those between their pretrained counterparts. 
For most model pairs, similarity drops to between 0.31 and 0.54, with an average of 0.46. 
Only a few pairs retain relatively high similarity.
This clearly shows that training increases representation similarity.

Overall, this comparison suggests that the high similarity between pretrained models cannot be explained by CKA estimation bias or model architecture alone. Instead, it is more likely related to the training process. This observation is consistent with recent findings on representation convergence~\cite{huh2024platonic,usama2026convergence}, where models trained on similar data distributions, objectives, and semantic tasks may learn similar representations.
Therefore, high representation similarity is not necessarily specific to model provenance.

\begin{figure*}[t]
  \centering
  \includegraphics[width=\linewidth]{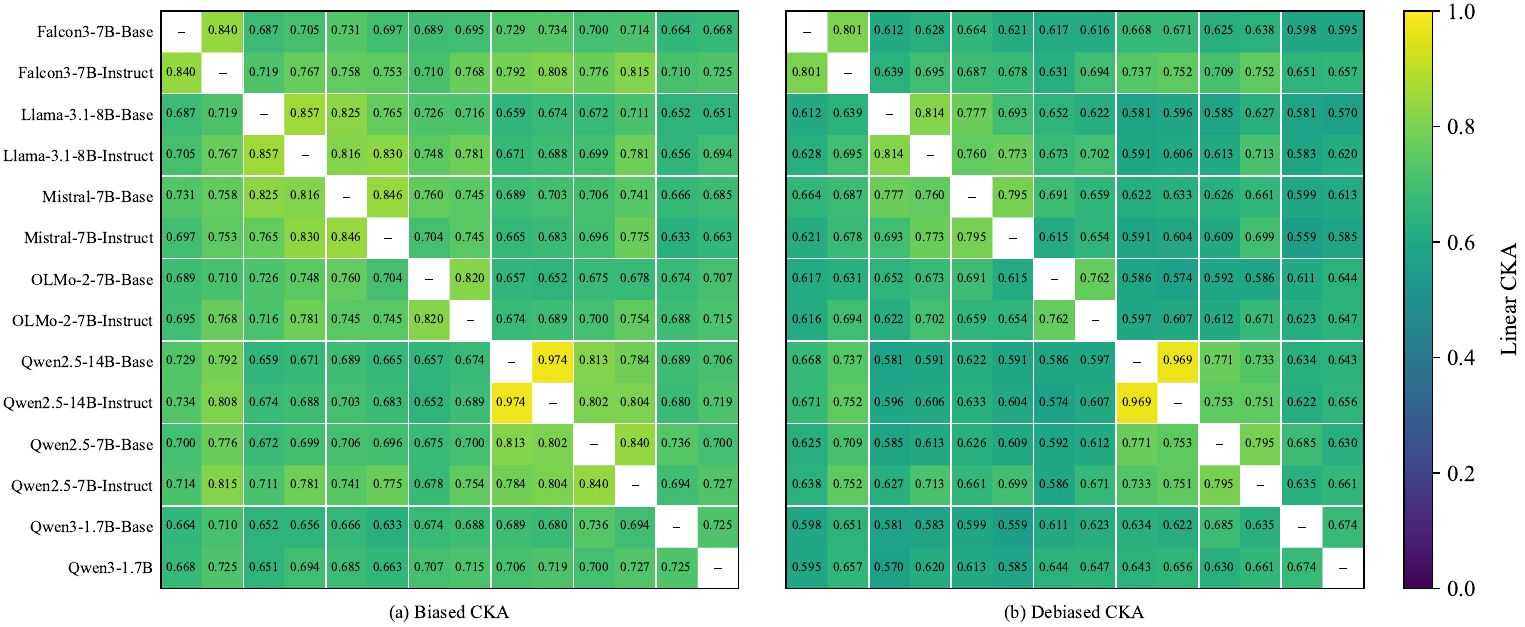}
  \vspace{-20pt}
  \caption{Pairwise similarity scores of REEF across different LLMs. Rows follow the same model order as the columns.}
  \label{fig4:REEF}
  \vspace{-10pt}
\end{figure*}

\subsection{Limitations of Representation Similarity for Model Fingerprinting}
The findings above suggest that representation similarity may not provide the separation between related and unrelated models required for fingerprint uniqueness.  We next examine this issue in existing fingerprinting methods.

Specifically, we analyze REEF~\cite{Zhang2025REEF}, which directly uses CKA between intermediate representations for provenance verification. For evaluation, the Base and Instruct versions of the seven model families introduced in Section~III-B are selected, forming a pool of 14 models.
The Base and Instruct models from the same family are treated as related pairs, while all other pairs are treated as unrelated. 
Following the original REEF setting, we use 200 probe samples from TruthfulQA and extract the hidden state of the last token at layer 18. 
We then compute pairwise similarity using both biased and debiased CKA.
Fig.~\ref{fig4:REEF}(a) shows the results with biased CKA. 
The entire matrix exhibits generally high similarity. 
Unrelated pairs have an average similarity of 0.718, and all scores exceed 0.5. Related pairs have a higher average similarity of 0.843, but the two distributions overlap. 
After debiasing, the average similarity of unrelated pairs decreases to 0.646, while that of related pairs decreases to 0.801, as shown in Fig.~\ref{fig4:REEF}(b).
However, the two distributions still overlap. 
For example, Qwen3-1.7b-Base is more similar to the unrelated Qwen2.5-7b-Base at 0.685 than to its own Instruct version at 0.674.

These results show that correcting estimation bias improves the reliability of similarity measurement but does not ensure fingerprint uniqueness. Unrelated models can still obtain similarity scores comparable to those of related models, because high representation similarity may arise from either model derivation or representation convergence. Therefore, representation similarity alone provides limited provenance discrimination. This motivates us to look beyond representations and examine how representations evolve across network depth.

\section{Computational Trajectory Analysis}
The previous section shows that unrelated LLMs can exhibit similar representations, limiting the ability of representations to distinguish model provenance. However, similar representations do not necessarily imply that they are formed through the same layer-wise evolution process. 
Each Transformer layer updates the current representation according to its parameters, and these successive updates form a computational trajectory. Unlike the representation at a single layer, the computational trajectory captures how the representation evolves through the network and retains provenance information specific to each model.

If computational trajectories contain provenance information, they should be discriminative across unrelated models while remaining stable under model derivation. To examine this, we visualize the computational trajectories of 56 models from seven LLM families using the same probe set. These models cover Base, Instruct, Fine-tuned~\cite{Chiang2023Vicuna}, Adapter-based~\cite{houlsby2019parameter}, Merged~\cite{yadav2023ties}, GPTQ-quantized~\cite{frantar2022gptq}, and Distilled~\cite{hinton2015distilling} variants. 
Since hidden spaces are not aligned across models, we represent each layer by the relative arrangement of its probe displacements. We then jointly project all layers into two dimensions using UMAP~\cite{mcinnes2018umap} and connect them by depth to visualize each model's computational trajectories.

As shown in Fig.~\ref{fig0_motivation}, models from different families generally exhibit distinct computational trajectories, whereas derived models tend to preserve the trajectories of their source models. These observations suggest that computational trajectories provide two desirable properties for provenance verification: uniqueness across unrelated models and robustness across derived models. Motivated by this, we use them as the basis for constructing TrFP.

\section{Trajectory Fingerprinting}
\begin{figure*}[t]
  \centering
  \includegraphics[width=\linewidth]{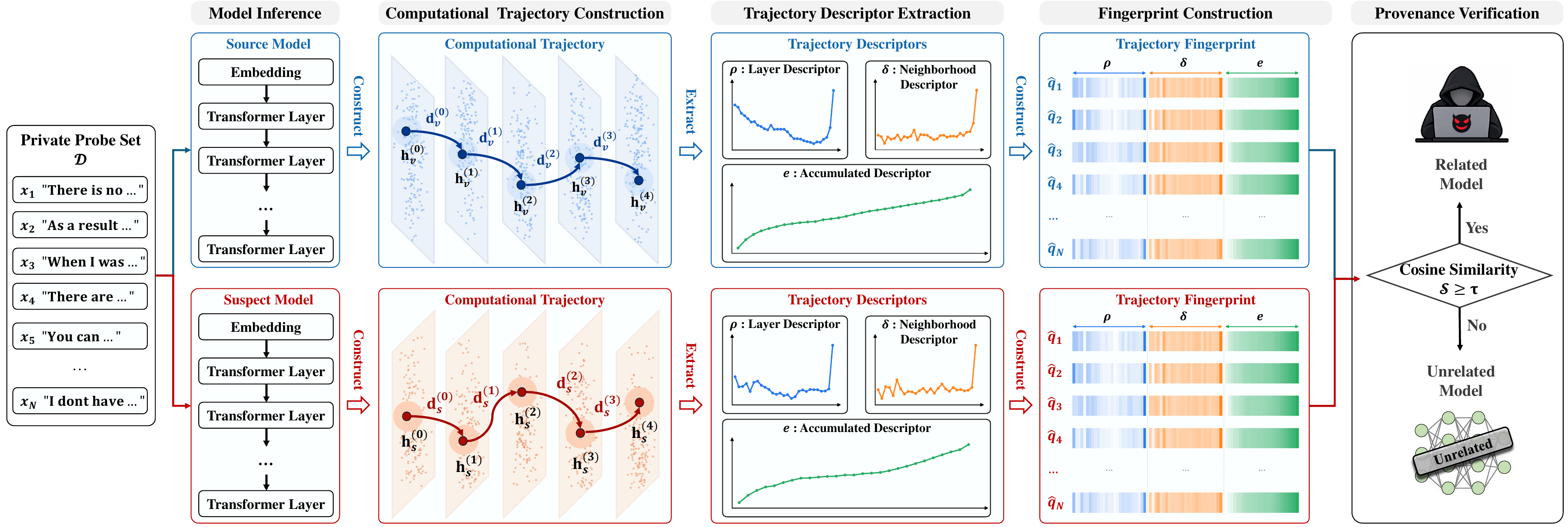}
     \vspace{-20pt}
  \caption{Overview of TrFP, which characterizes the layer-wise evolution of representations to construct compact fingerprints for model provenance verification.}
  \label{fig5:Overview}
  \vspace{-10pt}
\end{figure*}

The overall framework of TrFP is illustrated in Fig.~\ref{fig5:Overview}.
Given a private probe set and a model, TrFP first extracts its layer-wise hidden states and constructs the computational trajectories from their evolution across network depth.
It then extracts permutation-invariant trajectory descriptors and standardizes them to construct a compact trajectory fingerprint.
For provenance verification, TrFP determines the provenance relationship between the source and suspect models based on their fingerprint similarity.
We next describe each component in detail.

\subsection{Threat Model}
We study white-box model provenance verification for LLMs. Given a protected source model \(M_v\) and a suspect model \(M_s\), the objective is to determine whether \(M_s\) is related to \(M_v\). The pair is considered related if \(M_s\) is identical to or derived from \(M_v\) through model transformations. Otherwise, the pair is considered unrelated, even if the two models share the same architecture.

\textbf{Defender capabilities.}
We assume that the defender has white-box access to the source model \(M_v\) and can store its reference fingerprint in advance. During provenance verification, the defender has access to the intermediate hidden states of the suspect model \(M_s\), but does not require its training data or development history. Both models are evaluated using the same private probe set, allowing their computational trajectories to be extracted and compared.

\textbf{Attacker capabilities.}
We consider a strong adaptive attacker who obtains the source model and knows the fingerprinting algorithm used by the defender. The attacker may apply model transformations to obscure the provenance relationship while preserving the model's main functionality. Such transformations include fine-tuning, pruning, model merging, quantization, and parameter permutation. The attacker may also explicitly optimize the suspect model to reduce its fingerprint similarity to the source model. However, the defender's private probe set remains unknown to the attacker.

\subsection{Computational Trajectory Construction}
Let \(\mathcal{D}=\{x_i\}_{i=1}^{N}\) denote a private probe set with \(N\) probe samples, and let \(M\) denote a model with \(L\) Transformer layers.
We extract the hidden state of the last token at each layer, since under causal attention it aggregates information from the entire input~\cite{vaswani2017attention}.
The hidden-state sequence of $x_i$ across network depth is:
\begin{equation}
\left(\mathbf{h}_i^{(0)},\mathbf{h}_i^{(1)},\ldots,\mathbf{h}_i^{(L)}\right)=M(x_i),
\end{equation}
where $\mathbf{h}_i^{(0)}$ is the initial representation after the embedding layer, and $\mathbf{h}_i^{(l)}$ is the representation after the first $l$ Transformer layers.
To capture how representations change between consecutive layers, we define the difference between adjacent hidden states as the layer-wise displacement:
\begin{equation}
    \mathbf{d}_i^{(l)}=\mathbf{h}_i^{(l+1)}-\mathbf{h}_i^{(l)},\qquad l=0,\ldots,L-1,
\label{eq:disp}
\end{equation}
where $\mathbf{d}_i^{(l)}$ represents the displacement from $\mathbf{h}_i^{(l)}$ to $\mathbf{h}_i^{(l+1)}$.
The output of layer $l$ is determined by the current hidden state and the transformation defined by its parameters:
\begin{equation}
\mathbf{h}_i^{(l+1)}=\Phi_l\left(\mathbf{h}_i^{(l)};\boldsymbol{\theta}_l\right),
\label{eq:layer}
\end{equation}
where $\Phi_l(\cdot;\boldsymbol{\theta}_l)$ denotes the transformation defined by the parameters $\boldsymbol{\theta}_l$ of layer $l$. Therefore,
\begin{equation}
\mathbf{d}_i^{(l)}=\Phi_l\left(\mathbf{h}_i^{(l)};\boldsymbol{\theta}_l\right)-\mathbf{h}_i^{(l)}.
\label{eq:disp_phi}
\end{equation}
This shows that the layer-wise displacement $\mathbf{d}_i^{(l)}$ is determined by the current representation and the transformation defined by the layer parameters.
We define the computational trajectory $\mathcal{T}_M(x_i)$ of $x_i$ as the sequence of layer-wise displacements:
\begin{equation}
\mathcal{T}_M(x_i)
=
\left(
\mathbf{d}_i^{(0)},
\mathbf{d}_i^{(1)},
\ldots,
\mathbf{d}_i^{(L-1)}
\right).
\end{equation}

\subsection{Trajectory Descriptor Extraction}
Computational trajectories consist of high-dimensional displacements, which lie in the hidden space of each model.
As these spaces lack coordinate correspondence~\cite{raghu2017svcca,CKA}, trajectories cannot be compared directly across models.
Comparing them across models requires a compact representation that is comparable across hidden spaces and preserves the provenance information of the trajectory.
A computational trajectory is formed by successive layer-wise displacements along network depth, so we characterize it at three complementary levels. The layer level concerns the magnitude of each layer-wise displacement relative to the current representation. The neighborhood level concerns how this relative magnitude changes between neighboring layers. The accumulated level concerns the net displacement over all preceding layers. We construct one descriptor for each level.

At the layer level, we define the layer descriptor \(\rho_i^{(l)}\) as the relative update intensity of layer \(l\) for probe \(x_i\), namely the magnitude of its layer-wise displacement relative to its input representation:
\begin{equation}
\rho_i^{(l)}
=
\frac{\|\mathbf d_i^{(l)}\|_2}
{\|\mathbf h_i^{(l)}\|_2+\epsilon},
\label{eq:rho}
\end{equation}
where \(\epsilon\) ensures numerical stability. A larger \(\rho_i^{(l)}\) indicates that layer \(l\) alters its input representation more substantially.

At the neighborhood level, we define the neighborhood descriptor \(\delta_i^{(l)}\) as the change in relative update intensity of probe \(x_i\) from layer \(l-1\) to layer \(l\):
\begin{equation}
\delta_i^{(l)}
=
\begin{cases}
0, & l=0,\\
\rho_i^{(l)}-\rho_i^{(l-1)}, & l\geq1.
\end{cases}
\label{eq:delta}
\end{equation}
This change is positive when the relative update strengthens from layer \(l-1\) to layer \(l\), and negative when it weakens.

At the accumulated level, we define the accumulated descriptor \(e_i^{(l)}\) as the magnitude of the net displacement of probe \(x_i\) after layer \(l\), i.e., the vector sum of all preceding layer-wise displacements:
\begin{equation}
e_i^{(l)}
=
\left\|\sum_{k=0}^{l}\mathbf d_i^{(k)}\right\|_2
=
\left\|\mathbf h_i^{(l+1)}-\mathbf h_i^{(0)}\right\|_2.
\label{eq:e}
\end{equation}
Since the displacements are summed as vectors, \(e_i^{(l)}\) captures their directional interactions, becoming larger when they reinforce one another and smaller when they partially cancel.

Combining these three descriptors, the trajectory state of probe \(x_i\) at depth \(l\) is represented as:
\begin{equation}
\mathbf q_i^{(l)}
=
\left[
\rho_i^{(l)},
\delta_i^{(l)},
e_i^{(l)}
\right].
\end{equation}
All three descriptors depend only on Euclidean norms and thus require no correspondence between hidden dimensions. 
Together, they preserve the provenance information of the trajectory at the layer, neighborhood, and accumulated levels.

\subsection{Trajectory Fingerprint Construction}
The trajectory descriptors defined above characterize each probe at every network depth, but two issues arise when they are used directly for fingerprint construction. 
First, the three descriptors have different numerical scales, causing descriptors with larger magnitudes to dominate similarity computation. 
Second, at each network depth, the descriptors of different probes may share a baseline. When this baseline dominates, it obscures the variations across probes and weakens provenance discrimination.
Therefore, we standardize the trajectory descriptors before constructing the fingerprint.

For each layer-wise position \(l\), we define the descriptor center \(\mathbf{r}^{(l)}\in\mathbb{R}^{3}\) as the mean trajectory descriptor over the probe set \(\mathcal{D}=\{x_i\}_{i=1}^{N}\), which estimates the common baseline at this position:
\begin{equation}
\mathbf{r}^{(l)}=\frac{1}{N}\sum_{i=1}^{N}\mathbf{q}_i^{(l)},\qquad l=0,1,\ldots,L-1.
\end{equation}
We further use a global scale for each descriptor across all layer-wise positions. Specifically, we estimate the global scale \(\boldsymbol{\sigma}_M\in\mathbb{R}^{3}\) as:
\begin{equation}
\boldsymbol{\sigma}_M=\sqrt{\frac{1}{LN}\sum_{l=0}^{L-1}\sum_{i=1}^{N}\left(\mathbf{q}_i^{(l)}-\mathbf{r}^{(l)}\right)^2},
\end{equation}
where the square and square root are applied element by element. Based on the descriptor center $\mathbf{r}^{(l)}$ and global scale $\boldsymbol{\sigma}_M$, we define the standardized trajectory descriptor $\hat{\mathbf{q}}_i^{(l)}$ of probe sample \(x_i\) at layer-wise position \(l\) as:
\begin{equation}
\hat{\mathbf{q}}_i^{(l)}
=
\frac{\mathbf{q}_i^{(l)}-\mathbf{r}^{(l)}}{\boldsymbol{\sigma}_M+\epsilon},
\end{equation}
where \(\epsilon\) is a small constant for numerical stability. Finally, we define the trajectory fingerprint \(F_M\in\mathbb{R}^{L\times N\times3}\) of model \(M\) as the collection of standardized trajectory descriptors over all layer-wise positions and probe samples:
\begin{equation}
F_M=
\left[
\hat{\mathbf{q}}_i^{(l)}
\right]_{\begin{subarray}{l}
l=0,\ldots,L-1\\
i=1,\ldots,N
\end{subarray}}.
\end{equation}
The three dimensions of \(F_M\) correspond to layer-wise positions, probe samples, and trajectory descriptors, respectively.
The permutation invariance of these trajectory descriptors is analyzed in Section V-F.

\subsection{Fingerprint Comparison and Provenance Verification}
Given the trajectory fingerprints of a source model $M_v$ and a suspect model $M_s$, denoted by $F_{M_v} \in \mathbb{R}^{L_v \times N \times 3}$ and $F_{M_s} \in \mathbb{R}^{L_s \times N \times 3}$, respectively, TrFP determines their provenance relationship by comparing their fingerprint similarity.
However, different models may have different network depths, so their trajectory fingerprints cannot always be directly compared at the same layer-wise positions.
Therefore, we align the fingerprints by relative network depth and then compare the fingerprints at the aligned positions.

Specifically, we set \(\bar{L}=\min(L_v,L_s)\) as the number of aligned depth positions and uniformly sample these positions over relative network depth.
For the $k$-th aligned position, the indices in the source and suspect fingerprints are:
\begin{equation}
\begin{aligned}
p_k^v
&=
\operatorname{round}\left(
\frac{k(L_v-1)}{\bar{L}-1}
\right),\\
p_k^s
&=
\operatorname{round}\left(
\frac{k(L_s-1)}{\bar{L}-1}
\right),
\end{aligned}
\end{equation}
where $k=0,\ldots,\bar{L}-1$.
At each aligned position, we compute the cosine similarity for each probe input between the two fingerprints and average the similarities over all probe samples:
\begin{equation}
\gamma_k=\frac{1}{N}\sum_{i=1}^{N}\cos\left(F_{M_v}[p_k^v,i,:],F_{M_s}[p_k^s,i,:]\right).
\end{equation}
Here, \(\gamma_k\) denotes the average similarity at the \(k\)-th aligned position.
The overall fingerprint similarity is obtained by averaging over all aligned positions:
\begin{equation}
S(F_{M_v},F_{M_s})=\frac{1}{\bar{L}}\sum_{k=0}^{\bar{L}-1}\gamma_k.
\end{equation}
The resulting similarity $S(F_{M_v},F_{M_s})$ is compared with the decision
threshold $\tau$. 
If $S(F_{M_v},F_{M_s})\geq\tau$, the suspect model $M_s$ is considered to have a provenance relationship with the source model $M_v$.
Otherwise, it is considered unrelated.
For consistent evaluation with prior fingerprinting
methods~\cite{Zhang2025REEF,tang2026cref}, we set $\tau=0.5$ in our method.

\subsection{Analysis of Permutation Invariance}
Parameter permutation is a particularly important attack against model fingerprinting because it can potentially evade fingerprint-based verification at very low cost~\cite{entezari2021role,Zeng2024HuRef}.
Such attacks require no model retraining and preserve model functionality.
Therefore, permutation invariance is an important requirement for reliable provenance verification.
We next theoretically analyze the permutation invariance of TrFP under three common parameter permutation attacks.
Each layer \(\Phi_l\) in Eq.~\eqref{eq:layer} consists of a self-attention module and a feed-forward network (FFN) module.
For probe \(x_i\), let \(\mathbf{H}_i^{(l)}\) denote the hidden states of all its tokens at the input of the attention module in layer \(l\), whose last row is \(\mathbf h_i^{(l)}\). 
Let \(\mathbf{Z}_i^{(l)}\) denote the input of the FFN module in layer \(l\), obtained after \(\mathbf{H}_i^{(l)}\) passes through the attention module.

\textbf{1) Attention permutation.}
Consider a standard self-attention module. The query $\mathbf{Q}$, key $\mathbf{K}$, and value $\mathbf{V}$ are:
\begin{equation}
\mathbf{Q} = \mathbf{H}_i^{(l)}\mathbf{W}_Q, \quad
\mathbf{K} = \mathbf{H}_i^{(l)}\mathbf{W}_K, \quad
\mathbf{V} = \mathbf{H}_i^{(l)}\mathbf{W}_V,
\end{equation}
where \(\mathbf{W}_Q,\mathbf{W}_K,\mathbf{W}_V\) denote weights. The module output is
\begin{equation}
\operatorname{Attn}(\mathbf{H}_i^{(l)})
=
\operatorname{softmax}\left(
\frac{\mathbf{Q}\mathbf{K}^{\top}}{\sqrt{d_k}}
\right)
\mathbf{V}\mathbf{W}_O,
\end{equation}
where \(d_k\) is the attention dimension and \(\mathbf{W}_O\) is the output projection matrix.
Let \(\mathbf{P}_q\) and \(\mathbf{P}_v\) be permutation matrices applied consistently to the query/key and value dimensions, respectively. The corresponding transformations are:
\begin{equation}
\begin{aligned}
\widetilde{\mathbf{Q}} &=  \mathbf{H}_i^{(l)}\mathbf{W}_Q\mathbf{P}_q, \quad
\widetilde{\mathbf{K}} = \mathbf{H}_i^{(l)}\mathbf{W}_K\mathbf{P}_q,\\
\widetilde{\mathbf{V}} &= \mathbf{H}_i^{(l)}\mathbf{W}_V\mathbf{P}_v, \quad
\widetilde{\mathbf{W}}_O = \mathbf{P}_v^{-1}\mathbf{W}_O.
\end{aligned}
\end{equation}
Since permutation matrices are orthogonal,
\begin{equation}
\begin{aligned}
\widetilde{\operatorname{Attn}}(\mathbf{H}_i^{(l)})
&=
\operatorname{softmax}\left(
\frac{
(\mathbf{H}_i^{(l)}\mathbf{W}_Q\mathbf{P}_q)
(\mathbf{H}_i^{(l)}\mathbf{W}_K\mathbf{P}_q)^{\top}
}{\sqrt{d_k}}
\right)\\
&\quad
(\mathbf{H}_i^{(l)}\mathbf{W}_V\mathbf{P}_v)
(\mathbf{P}_v^{-1}\mathbf{W}_O)
=
\operatorname{Attn}(\mathbf{H}_i^{(l)}).
\end{aligned}
\end{equation}
Thus, the attention output, and hence \(\Phi_l\), remains unchanged. By Eq.~\eqref{eq:disp_phi}, the displacement \(\mathbf d_i^{(l)}\) and the trajectory descriptors $\rho_i^{(l)}$, $\delta_i^{(l)}$, $e_i^{(l)}$ also remain unchanged.

\textbf{2) FFN permutation.}
Consider a standard FFN:
\begin{equation}
\operatorname{FFN}(\mathbf{Z}_i^{(l)})=\sigma(\mathbf{Z}_i^{(l)}\mathbf{W}_1+\mathbf{b}_1)\mathbf{W}_2+\mathbf{b}_2,
\end{equation}
where \(\sigma(\cdot)\) denotes an element-wise activation function, and \(\mathbf{W}_1,\mathbf{W}_2,\mathbf{b}_1,\mathbf{b}_2\) denote weights and biases, respectively. Let $\mathbf{P}_f$ be a permutation matrix acting on the FFN intermediate dimension. A parameter permutation can be defined as:
\begin{equation}
\widetilde{\mathbf{W}}_1=\mathbf{W}_1\mathbf{P}_f, \quad
\widetilde{\mathbf{b}}_1=\mathbf{b}_1\mathbf{P}_f, \quad
\widetilde{\mathbf{W}}_2=\mathbf{P}_f^{-1}\mathbf{W}_2.
\end{equation}
Since the element-wise activation commutes with permutation, \(\mathbf{P}_f\) is canceled by \(\mathbf{P}_f^{-1}\) in the second linear projection:
\begin{equation}
\begin{aligned}
\widetilde{\operatorname{FFN}}(\mathbf{Z}_i^{(l)})
&=
\sigma\!\left(\mathbf{Z}_i^{(l)}\mathbf W_1\mathbf P_f+\mathbf b_1\mathbf P_f\right)
\mathbf P_f^{-1}\mathbf W_2+\mathbf b_2 \\
&=
\sigma\!\left(\mathbf{Z}_i^{(l)}\mathbf W_1+\mathbf b_1\right)
\mathbf P_f\mathbf P_f^{-1}\mathbf W_2+\mathbf b_2 \\
&=
\operatorname{FFN}(\mathbf{Z}_i^{(l)}).
\end{aligned}
\end{equation}
Thus, the FFN output, and hence \(\Phi_l\), remains unchanged. By Eq.~\eqref{eq:disp_phi}, the displacement \(\mathbf d_i^{(l)}\) and the trajectory descriptors $\rho_i^{(l)}$, $\delta_i^{(l)}$, $e_i^{(l)}$ also remain unchanged.

\textbf{3) Global hidden-dimension permutation.}
Let $\mathbf{P}_g$ be a permutation matrix acting on the hidden dimension.
All parameters involving the hidden dimension are permuted consistently, preserving the model function.
The hidden state at layer $l$ becomes:
\begin{equation}
\widetilde{\mathbf{h}}_i^{(l)}
=
\mathbf{h}_i^{(l)}\mathbf{P}_g.
\end{equation}
The layer-wise displacement in Eq.~\eqref{eq:disp} becomes:
\begin{equation}
\widetilde{\mathbf{d}}_i^{(l)}
=
\widetilde{\mathbf{h}}_i^{(l+1)}
-
\widetilde{\mathbf{h}}_i^{(l)}
=
\left(
\mathbf h_i^{(l+1)}
-
\mathbf h_i^{(l)}
\right)\mathbf{P}_g
=
\mathbf{d}_i^{(l)}\mathbf{P}_g.
\end{equation}
Since \(\mathbf P_g\) is orthogonal, the layer descriptor in Eq.~\eqref{eq:rho} remains unchanged:
\begin{equation}
\widetilde{\rho}_i^{(l)}
=
\frac{
\left\|
\widetilde{\mathbf d}_i^{(l)}
\right\|_2
}{
\left\|
\widetilde{\mathbf h}_i^{(l)}
\right\|_2+\epsilon
}
=
\frac{
\left\|
\mathbf d_i^{(l)}\mathbf{P}_g
\right\|_2
}{
\left\|
\mathbf h_i^{(l)}\mathbf{P}_g
\right\|_2+\epsilon
}
=
\rho_i^{(l)}.
\end{equation}
Since the neighborhood descriptor in Eq.~\eqref{eq:delta} is the difference between consecutive layer descriptors, it also remains unchanged, i.e., \( \widetilde{\delta}_i^{(l)} = \delta_i^{(l)}. \)
For the accumulated descriptor in Eq.~\eqref{eq:e}, orthogonality similarly gives:
\begin{equation}
\widetilde{e}_i^{(l)}
=
\left\|
\sum_{k=0}^{l}
\widetilde{\mathbf d}_i^{(k)}
\right\|_2
=
\left\|
\left(
\sum_{k=0}^{l}
\mathbf d_i^{(k)}
\right)\mathbf P_g
\right\|_2
=
e_i^{(l)}.
\end{equation}
In summary, parameter permutations either leave the computational trajectory unchanged or preserve its trajectory descriptors. Therefore, TrFP is permutation-invariant by construction.

\section{Experiments}
\subsection{Experimental Setup}
\textbf{Baselines and Implementation Details.}
We compare TrFP with five representative fingerprinting methods.
PCS~\cite{Zeng2024HuRef} and ICS~\cite{Zeng2024HuRef} are parameter-based methods, while REEF~\cite{Zhang2025REEF}, CREF~\cite{tang2026cref}, and AttnDiff~\cite{zhang2026attndiff} are representation-based methods.
For a fair comparison, we follow the original settings of each baseline whenever possible.
For PCS, we align corresponding parameters and compute their cosine similarity.
For ICS, we select 4,096 low-frequency tokens and use the input embeddings and the corresponding weights from the last two Transformer layers.
For REEF, we use 200 probe samples from TruthfulQA~\cite{Lin2022TruthfulQA} and extract the last-token hidden representation at layer 18.
For CREF, we construct 10 concept probe sets with 200 samples each.
For AttnDiff, we construct 60 pairs of semantically perturbed samples.
REEF, CREF, and AttnDiff use linear CKA 
to compute model similarity, while PCS, ICS, and TrFP use cosine similarity.
For TrFP, we use 200 probe samples from ToxiGen~\cite{hartvigsen2022toxigen}.

\textbf{Evaluation Metrics.}
We use metrics according to the objective of each experiment. 
For overall provenance verification, we report the mean similarities of related and unrelated model pairs, together with TPR and TNR at the fixed threshold \(\tau=0.5\). 
The unrelated mean and TNR characterize fingerprint uniqueness, while the related mean and TPR reflect the ability to identify related models. We further evaluate fingerprint robustness under specific model transformations, parameter permutation, and adaptive attacks.
For adaptive attacks, we additionally evaluate model utility, while storage cost is measured by the fingerprint size.
Following prior works~\cite{Zhang2025REEF,tang2026cref}, we regard a model pair as related if its similarity is not lower than \(0.5\), and as unrelated otherwise.

\begin{figure*}[t]
  \centering
  \includegraphics[width=\linewidth]{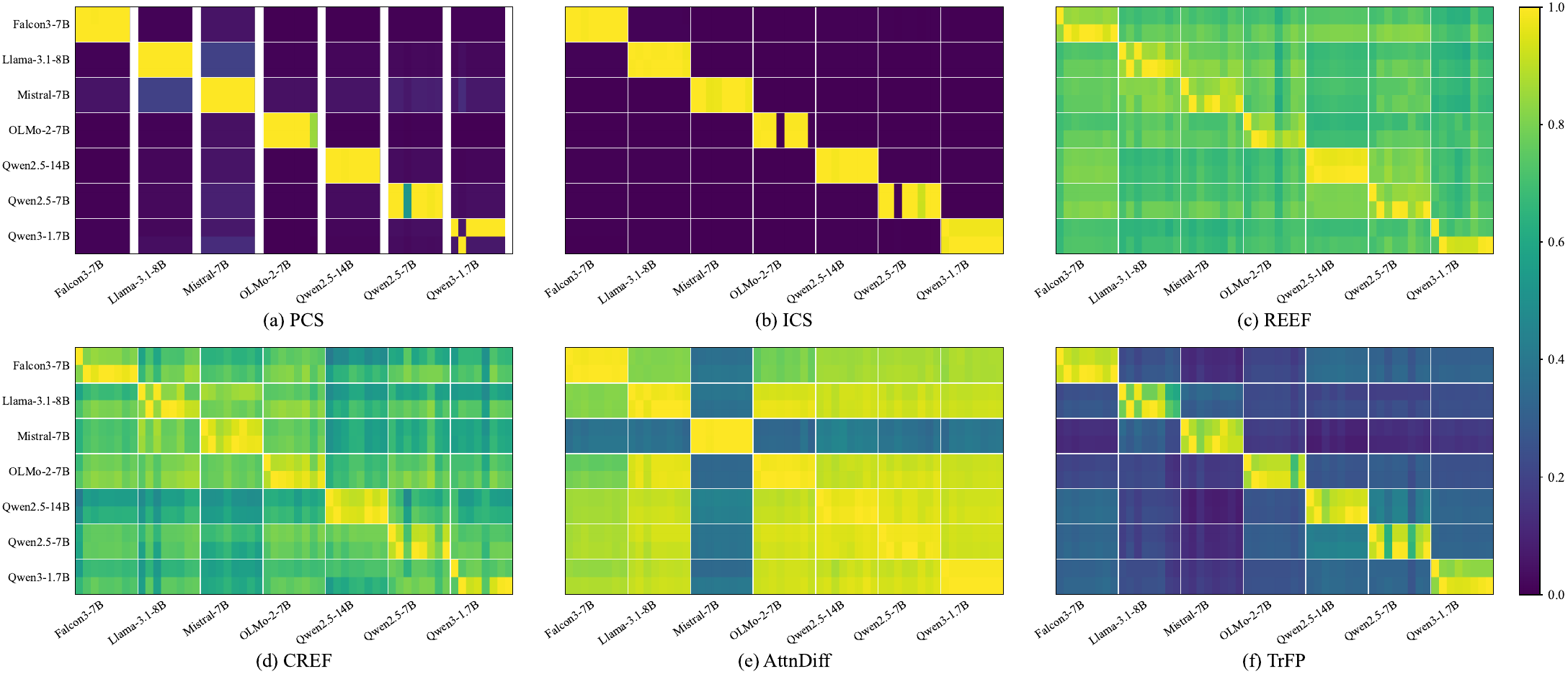}
     \vspace{-20pt}
  \caption{Pairwise similarity matrices of different fingerprinting methods. Rows correspond to the 14 source models and columns to the candidate models, both grouped by model family. The GPTQ-quantized models are excluded for PCS.}
  \label{fig6:heatmap}
  \vspace{-10pt}
\end{figure*}

\subsection{Overall Provenance Verification}
We select seven model families: Falcon3-7b~\cite{Falcon3}, Llama-3.1-8b~\cite{dubey2024llama3}, Mistral-7b~\cite{jiang2023mistral}, OLMo-2-7b~\cite{olmo2024olmo2}, Qwen2.5-14b~\cite{yang2024qwen25}, Qwen2.5-7b~\cite{yang2024qwen25}, and Qwen3-1.7b~\cite{yang2025qwen3}.
Each family contains eight candidate models, including Base, Instruct, Fine-tuned~\cite{Chiang2023Vicuna}, Adapter-based~\cite{houlsby2019parameter}, Merged~\cite{yadav2023ties}, GPTQ-quantized~\cite{frantar2022gptq}, and Distilled~\cite{hinton2015distilling} variants. 
This results in a candidate pool of 56 models.
Since downstream models are commonly derived from either Base or Instruct models, we use both versions from each family as source models, resulting in 14 source models.
After excluding self-comparisons, all methods except PCS produce a $14\times56$ similarity matrix, containing 98 related pairs and 672 unrelated pairs. 
Since PCS cannot compare GPTQ-quantized models, we exclude these seven models and obtain a $14\times49$ similarity matrix, containing 84 related pairs and 588 unrelated pairs.

Fig.~\ref{fig6:heatmap} shows the similarity matrices of the six methods. 
Ideally, related pairs should exhibit high similarity, forming diagonal blocks, while unrelated pairs outside these blocks should remain low. PCS, ICS, and TrFP all exhibit this desired block structure. In contrast, REEF, CREF, and AttnDiff show elevated similarity for many unrelated pairs.
Table~\ref{tab:overall_performance} further reports the quantitative results. 
PCS and ICS achieve strong separation between related and unrelated models, but still fail to identify all related pairs at the fixed threshold of 0.5, with TPR@0.5 values of 91.7\% and 93.9\%, respectively.
As representation-based baselines, REEF, CREF, and AttnDiff exhibit a different limitation. Although they obtain high similarities for related models, their unrelated mean similarities remain high at 0.728, 0.692, and 0.751, leading to TNR@0.5 values of only 0\%, 2.1\%, and 28.6\%, respectively. This indicates limited fingerprint uniqueness.
In comparison, TrFP achieves a related mean similarity of 0.887 and an unrelated mean similarity of 0.248, while attaining both 100\% TPR@0.5 and 100\% TNR@0.5.
These results show that TrFP reliably identifies related models while maintaining strong uniqueness.

\begin{table}[t]
\centering
\caption{Overall Provenance Verification Performance. }
\label{tab:overall_performance}

\renewcommand{\arraystretch}{1.05}

\begin{tabular*}{\columnwidth}{@{\extracolsep{\fill}}lcccc@{}}
\toprule
Method 
& Rel. Mean $\uparrow$
& Unrel. Mean $\downarrow$
& TPR@0.5 $\uparrow$
& TNR@0.5 $\uparrow$ \\
\midrule

PCS 
& \cellcolor{highgreen}0.906
& \cellcolor{highgreen}0.034
& \cellcolor{highgreen}91.7\%
& \cellcolor{highgreen}\textbf{100\%} \\

ICS 
& \cellcolor{highgreen}0.933
& \cellcolor{highgreen}\textbf{0.003}
& \cellcolor{highgreen}93.9\%
& \cellcolor{highgreen}\textbf{100\%} \\

REEF 
& \cellcolor{highgreen}0.889
& \cellcolor{lowred}0.728
& \cellcolor{highgreen}\textbf{100\%}
& \cellcolor{lowred}0\% \\

CREF 
& \cellcolor{highgreen}0.898
& \cellcolor{lowred}0.692
& \cellcolor{highgreen}\textbf{100\%}
& \cellcolor{lowred}2.1\% \\

AttnDiff 
& \cellcolor{highgreen}\textbf{0.988}
& \cellcolor{lowred}0.751
& \cellcolor{highgreen}\textbf{100\%}
& \cellcolor{lowred}28.6\% \\

\textbf{TrFP} 
& \cellcolor{highgreen}0.887
& \cellcolor{highgreen}0.248
& \cellcolor{highgreen}\textbf{100\%}
& \cellcolor{highgreen}\textbf{100\%} \\

\bottomrule
\end{tabular*}
   \vspace{-10pt}
\end{table}

\begin{table*}[t]
\centering
\caption{Fingerprint similarity under different model transformations. Green indicates similarity $\geq 0.8$, purple indicates $0.5 \leq$ similarity $< 0.8$, and red indicates similarity $<0.5$.}
\label{tab2:robust}
\resizebox{\textwidth}{!}{
\begin{tabular}{lcccccc}
\toprule

\multicolumn{7}{c}{\textbf{Model Fine-tuning}} \\
\cmidrule(lr){2-7}
& \makecell{Llama-2-finance-7b\\(5M Tokens)}
& \makecell{Vicuna-1.5-7b\\(370M Tokens)}
& \makecell{Wizardmath-7b\\(1.8B Tokens)}
& \makecell{Chinese-llama-2-7b\\(13B Tokens)}
& \makecell{Codellama-7b\\(500B Tokens)}
& \makecell{Llemma-7b\\(700B Tokens)} \\
\midrule

PCS       & \scorecell{0.9980} & \scorecell{0.9986} & \scorecell{0.9996} & \scorecell{0.9276} & \scorecell{0.5333} & \scorecell{0.5081} \\
ICS       & \scorecell{0.9943} & \scorecell{0.9941} & \scorecell{0.9993} & \scorecell{0.4626} & \scorecell{0.0175} & \scorecell{0.0159} \\
REEF      & \scorecell{0.9978} & \scorecell{0.9997} & \scorecell{0.9998} & \scorecell{0.9993} & \scorecell{0.9950} & \scorecell{0.9979} \\
CREF      & \scorecell{0.9946} & \scorecell{0.9992} & \scorecell{0.9993} & \scorecell{0.9897} & \scorecell{0.9959} & \scorecell{0.9964} \\
AttnDiff  & \scorecell{0.9989} & \scorecell{0.9987} & \scorecell{0.9975} & \scorecell{0.9968} & \scorecell{0.9864} & \scorecell{0.9831} \\
TrFP & \scorecell{0.9996} & \scorecell{0.9998} & \scorecell{0.9999} & \scorecell{0.9995} & \scorecell{0.9993} & \scorecell{0.9995} \\

\midrule
\midrule

\multicolumn{7}{c}{\textbf{Model Pruning}} \\
\cmidrule(lr){2-7}

& \makecell{Sheared-llama-1.3b\\-pruned}
& \makecell{Sheared-llama-1.3b}
& \makecell{Sheared-llama-1.3b\\-Sharegpt}
& \makecell{Sheared-llama-2.7b\\-pruned}
& \makecell{Sheared-llama-2.7b}
& \makecell{Sheared-llama-2.7b\\-Sharegpt} \\
\midrule

PCS       & \scorecell{0.0233} & \scorecell{0.0223} & \scorecell{0.0223} & \scorecell{0.0299} & \scorecell{0.0281} & \scorecell{0.0281} \\
ICS       & \scorecell{0.4504} & \scorecell{0.2894} & \scorecell{0.2893} & \scorecell{0.5721} & \scorecell{0.4070} & \scorecell{0.4041} \\
REEF      & \scorecell{0.6742} & \scorecell{0.5726} & \scorecell{0.4235} & \scorecell{0.6310} & \scorecell{0.6741} & \scorecell{0.9992} \\
CREF      & \scorecell{0.7116} & \scorecell{0.9439} & \scorecell{0.6939} & \scorecell{0.7354} & \scorecell{0.9637} & \scorecell{0.9981} \\
AttnDiff  & \scorecell{0.9903} & \scorecell{0.9922} & \scorecell{0.9912} & \scorecell{0.9942} & \scorecell{0.9945} & \scorecell{0.9945} \\
TrFP & \scorecell{0.8840} & \scorecell{0.8854} & \scorecell{0.8846} & \scorecell{0.9095} & \scorecell{0.9098} & \scorecell{0.9996} \\

\midrule
\midrule

\multicolumn{4}{c}{\textbf{GPTQ Quantization}}
& \multicolumn{3}{c}{\textbf{Model Merging (Evollm-jp-7b)}} \\
\cmidrule(lr){2-4}
\cmidrule(lr){5-7}

& \makecell{Falcon3-7b}
& \makecell{OLMo-2-7b}
& \makecell{Qwen3-1.7b}
& \makecell{shisa-gamma-7b-v1}
& \makecell{Wizardmath-7b-1.1}
& \makecell{Abel-7b-002} \\
\midrule

PCS
& \cellcolor{lowred}-
& \cellcolor{lowred}-
& \cellcolor{lowred}-
& \scorecell{0.9993}
& \scorecell{0.9988}
& \scorecell{0.9988} \\

ICS
& \scorecell{0.9993}
& \scorecell{0.0070}
& \scorecell{0.9795}
& \scorecell{0.9543}
& \scorecell{0.9296}
& \scorecell{0.9294} \\

REEF
& \scorecell{0.8393}
& \scorecell{0.8544}
& \scorecell{0.7254}
& \scorecell{0.9252}
& \scorecell{0.9082}
& \scorecell{0.8865} \\

CREF
& \scorecell{0.8131}
& \scorecell{0.9111}
& \scorecell{0.7389}
& \scorecell{0.9525}
& \scorecell{0.9611}
& \scorecell{0.9680} \\

AttnDiff
& \scorecell{0.9882}
& \scorecell{0.9820}
& \scorecell{0.9894}
& \scorecell{0.9940}
& \scorecell{0.9955}
& \scorecell{0.9966} \\

TrFP
& \scorecell{0.9196}
& \scorecell{0.8924}
& \scorecell{0.8286}
& \scorecell{0.8002}
& \scorecell{0.8063}
& \scorecell{0.8100} \\

\bottomrule
\end{tabular}
}
   \vspace{-10pt}
\end{table*}

\subsection{Robustness to Model Transformations}
To evaluate robustness under common model transformations, we consider four representative settings: fine-tuning, pruning, GPTQ quantization, and model merging.
For each transformation, we select source models and the derived models obtained from them through this transformation, and use all methods to compute the fingerprint similarity between each source model and its derived models.
Since each derived model is related to its source model, a higher similarity indicates stronger robustness to the transformation, whereas a lower similarity indicates weaker robustness.

\textbf{Fine-tuning.}
Fine-tuning updates model parameters through continued training and is one of the most common modifications after model release. 
We use Llama-2-7b~\cite{Touvron2023Llama2} as the source model and compare it with six derived models obtained by further training Llama-2-7b on different tasks and data scales: Llama-2-finance-7b, Vicuna-1.5-7b~\cite{Chiang2023Vicuna}, WizardMath-7b~\cite{Luo2025WizardMath}, Chinese-Llama-2-7b~\cite{Cui2023ChineseLLaMA}, CodeLlama-7b~\cite{Roziere2023CodeLlama}, and Llemma-7b~\cite{Azerbayev2024Llemma}. 
The results are shown in Table~\ref{tab2:robust}, where the number of training tokens for each derived model is given in parentheses.  
PCS and ICS show clear similarity drops on some fine-tuned models, indicating that extensive fine-tuning may affect parameter-based fingerprints. 
In contrast, REEF, CREF, AttnDiff, and TrFP maintain high similarity overall. 
In particular, TrFP achieves similarity close to 1 across all fine-tuned models, showing strong robustness to fine-tuning.

\textbf{Pruning.}
Pruning compresses a model by removing model parameters or structures. 
We use Llama-2-7b~\cite{Touvron2023Llama2} as the source model and compare it with the 1.3b and 2.7b Sheared-Llama~\cite{Xia2024ShearedLLaMA} and their variants.
For each scale, the pruned variant is obtained directly from Llama-2-7b through structured pruning, Sheared-Llama is obtained by further pretraining the pruned variant, and the ShareGPT variant is obtained by further fine-tuning Sheared-Llama on ShareGPT instruction data.
The results are shown in Table~\ref{tab2:robust}.
PCS almost completely loses similarity after pruning, while ICS remains below 0.58 in all cases. REEF and CREF perform better but exhibit considerable variation across different derived models.
In contrast, AttnDiff and TrFP remain stable across all pruned models, indicating robustness to substantial structural changes.

\textbf{GPTQ Quantization.}
Quantization reduces model precision and changes the numerical values of model parameters while preserving model functionality.
We use Falcon3-7b, OLMo-2-7b, and Qwen3-1.7b as source models and compare each of them with its GPTQ-quantized version.
PCS is not applicable to GPTQ-quantized models because the quantized parameter representation is incompatible with direct parameter comparison.
As shown in Table~\ref{tab2:robust}, ICS remains robust on most models but shows instability on OLMo-2-7b.
REEF and CREF retain relatively high similarity after quantization, although some variation is observed across different model families.
AttnDiff remains highly stable across all models, while TrFP also maintains consistently high similarity.
Overall, TrFP shows strong robustness to reduced parameter precision.

\textbf{Model Merging.}
Model merging combines parameters from multiple source models, making provenance verification more challenging because the resulting model may inherit characteristics from several sources.
We use Evollm-jp-7b as the merged model and compare it with its three constituent models: shisa-gamma-7b-v1, Wizardmath-7b-1.1, and Abel-7b-002.
As shown in Table~\ref{tab2:robust}, most methods retain substantial similarity between the merged model and its constituent models.
Because model merging integrates parameters from multiple source models, its computational trajectory may partially deviate from that of each source model. As a result, TrFP yields slightly lower similarity scores than the other methods. Nevertheless, all similarities remain well above 0.5, indicating that TrFP can still reliably identify these provenance relationships.

\begin{table*}[t]
\centering
\caption{Robustness Under Parameter Permutation Attacks. Results Are Reported as Mean $\pm$ Standard Deviation Over Three Seeds}
\label{tab3:permutation}
\setlength{\tabcolsep}{5.0pt}
\renewcommand{\arraystretch}{1.05}

\begin{tabular}{lcccccccc}
\toprule
& \multicolumn{4}{c}{\textbf{Qwen2.5-7b}}
& \multicolumn{4}{c}{\textbf{Llama-3.1-8b}} \\
\cmidrule(lr){2-5}
\cmidrule(lr){6-9}

\textbf{Method}
& \textbf{FFN}
& \textbf{Attention}
& \textbf{Hidden}
& \textbf{Combined}
& \textbf{FFN}
& \textbf{Attention}
& \textbf{Hidden}
& \textbf{Combined} \\
\midrule

PCS
& 0.5895$\pm$0.0000
& 0.5551$\pm$0.0069
& 0.4744$\pm$0.0001
& 0.0941$\pm$0.0028
& 0.3716$\pm$0.0000
& 0.7753$\pm$0.0027
& 0.0427$\pm$0.0001
& 0.0426$\pm$0.0001 \\

ICS
& 1.0000$\pm$0.0000
& 1.0000$\pm$0.0000
& 1.0000$\pm$0.0000
& 1.0000$\pm$0.0000
& 1.0000$\pm$0.0000
& 1.0000$\pm$0.0000
& 1.0000$\pm$0.0000
& 1.0000$\pm$0.0000 \\

REEF
& 0.9998$\pm$0.0000
& 0.9998$\pm$0.0000
& 0.9998$\pm$0.0000
& 0.9998$\pm$0.0000
& 0.9999$\pm$0.0000
& 0.9999$\pm$0.0000
& 0.9999$\pm$0.0000
& 0.9998$\pm$0.0000 \\

CREF
& 0.9996$\pm$0.0001
& 0.9996$\pm$0.0000
& 0.9994$\pm$0.0001
& 0.9994$\pm$0.0001
& 0.9995$\pm$0.0002
& 0.9995$\pm$0.0000
& 0.9994$\pm$0.0001
& 0.9994$\pm$0.0000 \\

AttnDiff
& 0.9993$\pm$0.0001
& 0.9993$\pm$0.0001
& 0.9961$\pm$0.0007
& 0.9960$\pm$0.0012
& 1.0000$\pm$0.0001
& 0.9999$\pm$0.0000
& 0.9999$\pm$0.0000
& 0.9999$\pm$0.0001 \\

\textbf{TrFP}
& 0.9970$\pm$0.0002
& 0.9968$\pm$0.0002
& 0.9968$\pm$0.0000
& 0.9970$\pm$0.0003
& 0.9974$\pm$0.0002
& 0.9979$\pm$0.0002
& 0.9973$\pm$0.0002
& 0.9970$\pm$0.0004 \\

\bottomrule
\end{tabular}
\vspace{-10pt}
\end{table*}

\subsection{Robustness to Parameter Permutation Attacks}
As discussed in Section V-F, parameter permutation is an important robustness test for model fingerprinting.
Therefore, we evaluate all methods under four permutations: FFN permutation, attention permutation, hidden-dimension permutation, and a combined permutation of all three.
We use Qwen2.5-7b and Llama-3.1-8b as source models.
Each permutation is applied to all permutable units and repeated with three seeds.
All experiments use BF16 precision, which may introduce small deviations from theoretical permutation invariance.

The results are shown in Table~\ref{tab3:permutation}.
PCS suffers clear similarity degradation under parameter permutation, especially under hidden-dimension and combined permutations, because its direct parameter comparison relies on fixed parameter correspondence.
In contrast, REEF, CREF, and AttnDiff remain highly stable by using CKA-based representation comparison, which is insensitive to coordinate permutations.
ICS also remains stable because its fingerprint is constructed from permutation-invariant terms of the attention and FFN weights.
TrFP maintains high similarity under all four permutations, consistent with the theoretical analysis in Section~V-F.

\begin{figure}[t]
  \centering
  \includegraphics[width=\linewidth]{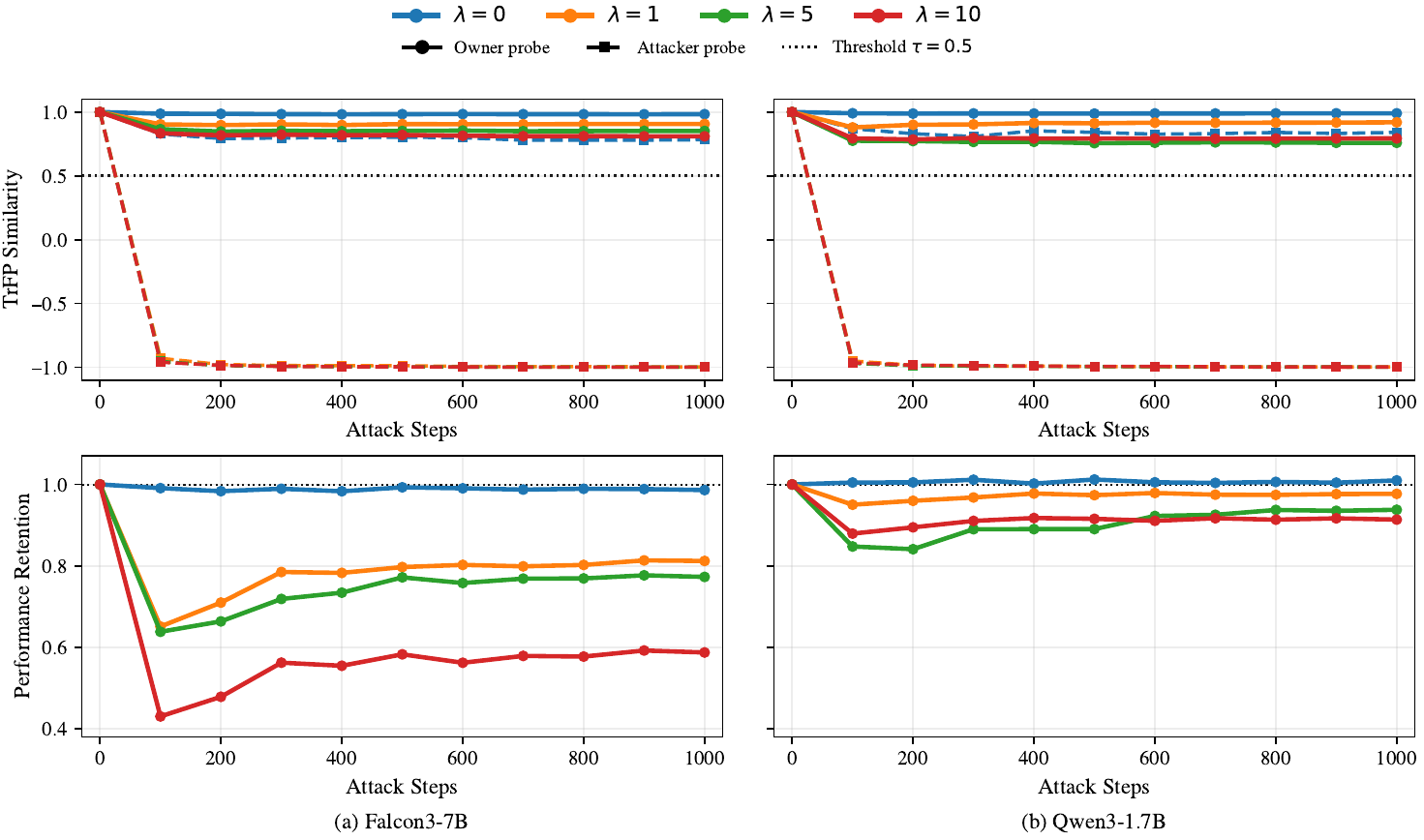}
    \vspace{-20pt}
  \caption{TrFP similarity and performance retention under adaptive attacks with different attack strengths $\lambda$. }
  \label{fig7:adaptive_attack}
  \vspace{-10pt}
\end{figure}

\subsection{Robustness to Adaptive Attacks}
The above experiments mainly consider model transformations that are independent of the fingerprinting mechanism. 
We further consider a stronger adaptive attack, where the attacker knows the complete fingerprint extraction and matching procedure of TrFP and directly optimizes the model to evade provenance verification~\cite{yao2023removalnet}. 
Following Kerckhoffs's principle, the design of TrFP is public, while the model owner's probe set is kept private.
Therefore, the attacker constructs a surrogate probe set to estimate the fingerprint and optimize the model accordingly.

Specifically, the attacker adds a fingerprint evasion loss to the standard task loss, forcing the computational trajectory of the fine-tuned model to deviate from that of the source model on the attacker's probe set. The attack objective is:
\begin{equation}
\mathcal{L}_{\mathrm{attack}}=\mathcal{L}_{\mathrm{task}}+\lambda \mathcal{L}_{\mathrm{evade}},
\end{equation}
where $\mathcal{L}_{\mathrm{task}}$ is the language modeling loss used to preserve model utility, and $\mathcal{L}_{\mathrm{evade}}$ is the TrFP similarity between the current model and the source model on the attacker's probe set.
The coefficient $\lambda$ controls the attack strength.
Minimizing this objective balances model utility and provenance evasion.

We use Falcon3-7b and Qwen3-1.7b as source models and perform adaptive fine-tuning with LoRA~\cite{hu2021lora}.
The attacker uses Alpaca~\cite{taori2023stanford} for fine-tuning and constructs a surrogate probe set using 200 samples from TruthfulQA.
The model owner uses 200 samples from ToxiGen as a private probe set.
We set $\lambda$ to 0, 1, 5, and 10 and optimize the model for 1,000 steps.
TrFP similarity on both probe sets and model performance are evaluated every 100 steps.

The results are shown in Fig.~\ref{fig7:adaptive_attack}.
For both models, the attack rapidly drives the similarity on the attacker's surrogate probes toward $-1$, showing that fingerprint similarity can be effectively suppressed when the probes are known.
However, the attack transfers poorly to the model owner's private probe set, where the similarity remains above the verification threshold $\tau=0.5$ throughout the attack.
Stronger attacks reduce the similarity on the private probe set only to a limited extent, and this reduction quickly saturates as $\lambda$ increases.
Meanwhile, larger $\lambda$ values cause increasingly severe utility degradation.
For Falcon3-7b, increasing $\lambda$ from 5 to 10 brings little additional reduction in fingerprint similarity on the private probes, while model performance deteriorates substantially.
Overall, these results show that TrFP remains robust to adaptive attacks, even when the fingerprinting mechanism is fully known.

\begin{table}[t]
\centering
\caption{Ablation Study of Trajectory Descriptors}
\label{tab:descriptor_ablation}
\renewcommand{\arraystretch}{1.05}

\resizebox{\columnwidth}{!}{
\begin{tabular}{lcccccc}
\toprule
& \multicolumn{2}{c}{\textbf{Effectiveness}}
& \multicolumn{4}{c}{\textbf{Robustness}} \\
\cmidrule(lr){2-3}
\cmidrule(lr){4-7}
Method
& TPR
& TNR
& Fine-tuning
& Pruning
& GPTQ
& Merging \\
\midrule

$\rho$
& 97.96\%
& 100.0\%
& 0.8327
& 0.7455
& 0.7648
& 0.6958 \\

$\delta$
& 92.86\%
& 100.0\%
& 0.5693
& 0.3122
& 0.7171
& 0.6288 \\

$e$
& 95.92\%
& 93.15\%
& 0.9587
& 0.8845
& 0.7917
& 0.7139 \\

$\rho+\delta$
& 97.96\%
& 100.0\%
& 0.8277
& 0.6662
& 0.8432
& 0.7736 \\

$\rho+e$
& 100.0\%
& 100.0\%
& 0.9603
& 0.8850
& 0.8412
& 0.7673 \\

$\delta+e$
& 98.98\%
& 100.0\%
& 0.9623
& 0.8860
& 0.8361
& 0.7507 \\

$\rho+\delta+e$
& \textbf{100.0\%}
& \textbf{100.0\%}
& \textbf{0.9996}
& \textbf{0.9121}
& \textbf{0.8803}
& \textbf{0.8055} \\

\bottomrule
\end{tabular}
}
\vspace{-10pt}
\end{table}

\begin{figure*}[t]
  \centering
  \includegraphics[width=\linewidth]{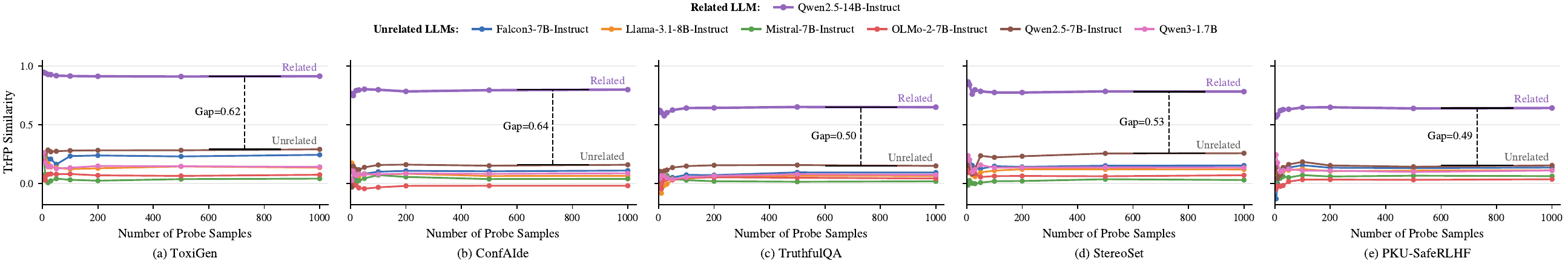}
      \vspace{-20pt}
  \caption{TrFP similarity under different probe set sizes on five probe datasets. Qwen2.5-14b is the source model, Qwen2.5-14b-Instruct is the related model, and the Instruct models from the other six families are unrelated models.}

  \label{fig8:datasetablation}
  \vspace{-10pt}
\end{figure*}

\begin{table*}[t]
\centering
\caption{Storage Cost (\textnormal{MiB}) of Different Fingerprinting Methods on Various LLMs}
\label{tab:storage_cost}
\setlength{\tabcolsep}{5pt}
\renewcommand{\arraystretch}{1.05}
\begin{tabular}{lrrrrrr}
\toprule
Method
& Llama2-7b
& Llama2-70b
& GLM-4.5-Air-106b
& GLM-4.5-355b
& DeepSeek-V3-671b
& KimiK2-1043b \\
\midrule
PCS
& 26702.8800
& 267028.8100
& 404357.9100
& 1354217.5300
& 2559661.8700
& 3978749.2500 \\

ICS
& 384.0000
& 384.0000
& 384.0000
& 384.0000
& 384.0000
& 384.0000 \\

REEF
& 3.1250
& 6.2500
& 3.1250
& 3.9063
& 5.4688
& 5.4688 \\

CREF
& 0.0076
& 0.0076
& 0.0076
& 0.0076
& 0.0076
& 0.0076 \\

AttnDiff
& 0.7031
& 0.5156
& 3.0322
& 6.0645
& 5.3613
& 2.6807 \\

\textbf{TrFP}
& 0.0732
& 0.1831
& 0.1053
& 0.2106
& 0.1396
& 0.1396 \\
\bottomrule
\end{tabular}
   \vspace{-10pt}
\end{table*}

\subsection{Ablation Study}

We conduct ablation studies from three aspects: trajectory descriptor composition, probe dataset, and probe set size. 

\textbf{Effect of Trajectory Descriptor Composition.}
To examine the contribution of each trajectory descriptor, we evaluate all seven non-empty combinations of $\rho$, $\delta$, and $e$, while keeping all other experimental settings unchanged.
The results are shown in Table~\ref{tab:descriptor_ablation}.
Among the individual descriptors, $\rho$ provides the strongest effectiveness, achieving 97.96\% TPR and 100\% TNR, while $e$ shows substantially stronger robustness to fine-tuning and pruning.
In contrast, $\delta$ alone is less effective, but combining it with $\rho$ notably improves robustness to GPTQ quantization and model merging.
Finally, the complete combination $\rho+\delta+e$ achieves 100\% TPR and TNR and the highest robustness under all four transformations.
These results show that the three descriptors capture complementary trajectory information. Therefore, we use $\rho+\delta+e$ as the default configuration.

\textbf{Effect of Probe Dataset.}
To examine whether TrFP depends on a specific type of probe data, we evaluate five datasets with different text characteristics: ToxiGen~\cite{hartvigsen2022toxigen}, ConfAIde~\cite{mireshghallah2024can},  TruthfulQA~\cite{Lin2022TruthfulQA}, StereoSet~\cite{nadeem2021stereoset}, and PKU-SafeRLHF~\cite{ji2024pku}, covering toxicity, privacy, factuality, bias, and safety, respectively. We use Qwen2.5-14b as the source model, Qwen2.5-14b-Instruct as the related model, and Instruct models from the other six model families as unrelated models. All other experimental settings remain the same as in the main experiment.
The results are shown in Fig.~\ref{fig8:datasetablation}. Although the absolute similarities vary across datasets, Qwen2.5-14b-Instruct consistently achieves substantially higher similarity than the unrelated models on all five datasets. This shows that the provenance discrimination of TrFP is not sensitive to the choice of probe dataset. In this paper, we use ToxiGen as the default probe dataset.

\textbf{Effect of Probe Set Size.}
We further examine the effect of the probe set size $N$ on TrFP.
The results are shown in Fig.~\ref{fig8:datasetablation}. When $N$ is between 10 and 20, the similarities show noticeable fluctuations, indicating that the trajectory fingerprint cannot be estimated reliably from very small probe sets. As $N$ increases, the results gradually stabilize. When $N$ reaches approximately 100 to 200, the results are already close to those obtained with $N=1000$, while further increasing $N$ brings only minor changes. These results show that TrFP can obtain stable provenance fingerprints without requiring a large probe set. Therefore, we set $N=200$ by default as a balance between stability and evaluation cost.

\subsection{Storage Cost}
In practical provenance verification, reference fingerprints must be stored for registered models. As the model registry grows, storage efficiency becomes increasingly important. Therefore, we compare the storage costs of different methods. For each method, we only count the final fingerprint required for verification, excluding the model itself and temporary variables used during fingerprint extraction.
Table~\ref{tab:storage_cost} reports the fingerprint sizes of different methods on Llama2-7b, Llama2-70b~\cite{Touvron2023Llama2}, GLM-4.5-Air-106b~\cite{Zeng2025GLM45}, GLM-4.5-355b~\cite{Zeng2025GLM45}, DeepSeek-V3-671b~\cite{Liu2024DeepSeekV3}, and KimiK2-1043b~\cite{Kimi2025K2}. PCS directly depends on model parameters, so its storage cost increases rapidly with model size. ICS compresses model parameters into invariant features and substantially reduces the storage cost, but still requires approximately 384 MiB. 
REEF and AttnDiff further reduce the storage cost to the MiB level, while CREF produces the most compact fingerprint at only 0.0076 MiB. 
TrFP also maintains a low storage cost, with fingerprint sizes changing only slightly as the model scale increases from 7b to 1043b. This is because it compresses high-dimensional computational trajectories into compact trajectory descriptors without storing model parameters or raw representations. Therefore, TrFP's strong provenance performance does not rely on larger fingerprints, making it well suited for large-scale model provenance verification.

\section{Conclusion}
In this paper, we propose TrFP, an intrinsic fingerprinting method for LLM provenance verification based on computational trajectories. 
Rather than relying on model parameters or representations, TrFP characterizes how representations evolve across network depth using compact and permutation-invariant trajectory descriptors.
This approach maintains robustness to model transformations while ensuring uniqueness against unrelated models.
Extensive experiments validate the effectiveness of TrFP, showing clear separation between related and unrelated models even under various model transformations, parameter permutation, and adaptive attacks.
Its permutation invariance and small storage cost make it practical for large-scale model provenance verification. 
However, TrFP currently requires access to intermediate representations, making it primarily applicable to white-box provenance verification. Its robustness across broader model architectures and stronger adaptive attacks remains to be further studied. 
Future work will investigate these settings and extend TrFP from LLMs to agentic systems by exploring provenance characteristics in planning, tool use, and interaction trajectories.

\bibliographystyle{IEEEtran}
\bibliography{sample-base}

\end{document}